\documentclass[usenatbib]{mn2e}
\usepackage{epsfig}
\usepackage{amsmath}
\usepackage{graphicx}
\usepackage{array}
\usepackage{textcomp}
\usepackage{amssymb}

\newcolumntype{L}[1]{>{\raggedright\let\newline\\\arraybackslash\hspace{0pt}}m{#1}}
\newcolumntype{C}[1]{>{\centering\let\newline\\\arraybackslash\hspace{0pt}}m{#1}}
\newcolumntype{R}[1]{>{\raggedleft\let\newline\\\arraybackslash\hspace{0pt}}m{#1}}

\title[BCG star formation as measured by WISE]
      {The Rarity of Star Formation in Brightest Cluster Galaxies as Measured by WISE}
            
\author[A.\ Fraser-McKelvie, M.\ J.\ I.\ Brown, \& K.\ A.\ Pimbblet]
       {Amelia Fraser-McKelvie$^{1,2}$\thanks{amelia.mckelvie@.monash.edu}, Michael J.\ I.\ Brown$^{1,2}$, Kevin A.\ Pimbblet$^{1,2,3}$
        \vspace*{1mm}\\
        $^{1}$ School of Physics, Monash University, Clayton, Victoria 3800, Australia\\
	$^{2}$ Monash Centre for Astrophysics (MoCA), Monash University, Clayton, Victoria 3800, Australia\\
	$^{3}$ Department of Physics and Mathematics, University of Hull, Cottingham Road, Kingston-upon-Hull, HU6 7RX, UK\\
	}

\begin{document}
 \maketitle
 \begin{abstract}
We present the mid-infrared (IR) star formation rates of 245 X-ray selected, nearby ($z<0.1$) brightest cluster galaxies (BCGs).
A homogeneous and volume limited sample of BCGs was created by X-ray selecting clusters with $L_{x}>1\times10^{44}~\textrm{erg}~\textrm{s}^{-1}$.
The Wide-Field Infrared Survey Explorer (WISE) AllWISE Data Release provides the first measurement of the 12 $ \mu$m star formation indicator for all BCGs in the nearby Universe.
Perseus A and Cygnus A are the only galaxies in our sample to have star formation rates of $> 40~\textrm{M}_{\odot}~ \textrm{yr}^{-1}$,  indicating that these two galaxies are highly unusual at current times.
Stellar populations of $99 \pm 0.6\%$ of local BCGs are (approximately) passively evolving, with star formation rates of $<10~\textrm{M}_{\odot}~ \textrm{yr}^{-1}$. We find that in general, star formation produces only modest BCG growth at the current epoch. 
 \end{abstract}
 
 \begin{keywords}
  galaxies: clusters: general -- galaxies: elliptical and lenticular, cD -- galaxies: star formation -- infrared: galaxies
 \end{keywords}

\section{Introduction}
%BCGS ARE RED AND DEAD
Brightest cluster galaxies (BCGs) are massive, highly luminous ellipticals located at the bottom of a galaxy cluster's potential well.
They differ from other large elliptical galaxies in size and velocity dispersion \citep[e.g.,][]{vonderLinden07} and are often offset from the cluster red sequence \citep{Bildfell08, Bernardi07}, 
indicating differences in stellar populations and assembly histories \citep[e.g.,][]{Brough07, Liu08}. 
In the local Universe, the bulk of BCGs are red, early-type galaxies with spectra lacking emission lines.
Any activity is largely quenched, likely due to a combination of dwindling 
merger activity and successful feedback mechanisms such as virial shock heating and AGN feedback \citep[e.g.,][and references therein]{Binney95,Voit05,Dekel06,DeLucia07}.

%WHILE MOST ARE RED, MANY HAVE A TRICKLE OF SF
Despite the predominantly quiescent population at current times, many BCGs possess trickles of star formation \citep[e.g.,][]{Bildfell08,Pipino09,Ford13}.
%Hoffer para
The central cluster environment is known to affect BCG star formation rates (SFRs), for example the presence of a cooling flow is likely to trigger star formation \citep[e.g.,][and references therein]{ODea08,Stott08,Hicks10,Donahue09,Rawle12}.

Cooling flow clusters, or clusters with low central gas entropy, were found to produce infrared (IR) and UV excess in the X-ray selected cluster sample of \citet{Hoffer12}.
Of the 243 BCGs in the \citet{Hoffer12} sample, BCGs located in low central gas entropy clusters had an excess above that expected for purely old stellar emission of 43\% and 38\% in the IR and UV respectively.
SFRs determined from the excess emission were moderate, and of all galaxies in the local Universe (defined as $z<$0.1), only Perseus A and Cygnus A possessed SFRs $>$10 $\textrm{M}_{\odot}~\textrm{yr}^{-1}$ (34 and 95 $\textrm{M}_{\odot}~\textrm{yr}^{-1}$ respectively). 
The \citet{Hoffer12} sample was heterogeneous but uniformly characterised, including only clusters that had been observed by Chandra. 

Optically derived BCG SFRs were provided for the X-ray selected sample of \citet{Crawford99}. Emission line spectra comprised 27\% of the BCG sample, and a further 6\% had only NII in emission and H$\alpha$ in absorption. 
After fitting a stellar spectra template to the galaxies in their sample with high H$\alpha$ luminosities, they found optically-derived SFRs of $<$1--10s of $\textrm{M}_{\odot}~\textrm{yr}^{-1}$. One cluster at a higher redshift (Abell 1835, $z$=0.253) had a BCG with a SFR of 125 $\textrm{M}_{\odot}~\textrm{yr}^{-1}$. 

%HERE ARE SOME BCGS WITH A LARGE SFR, SOME ARE AT HIGH Z
At $z>1$, cluster samples with star forming central galaxies are common \citep[e.g.,][]{Brodwin13}. At intermediate redshifts there are some examples of starburst BCGs, such as the Phoenix cluster (z=0.596) BCG which has a SFR of 740 $\textrm{M}_{\odot}~\textrm{yr}^{-1}$ \citep{McDonald12}.
In the local Universe, examples of star forming BCGs are rare, but Perseus A (NGC 1275) is an exception to the typically quiescent BCG population.
Displaying optical emission lines \citep[e.g.,][]{Fabian08}, a Seyfert 1 nucleus \citep[e.g.,][]{Burbidge65} and cold molecular gas near the core 
\citep[e.g.,][]{Bridges98,Fabian94,Edge03}, Perseus A 
%possesses a SFR of $\sim30~\textrm{M}_{\odot}~\textrm{yr}^{-1}$ \citep{Wirth83}, 
possesses a high SFR \citep{Wirth83}, despite significant AGN activity \citep[as demonstrated by][]{Fabian08}. 
In fact, Perseus A is the archetypal example of AGN feedback in the local Universe, despite its high SFR.

%BUT WE DONÕT KNOW HOW COMMON THIS IS
Whether the activity of Perseus A is unique or a common stage in a BCG's life cycle is unknown.
While it is clear that star forming BCGs do exist, and that a significant population of them can be found in environments where we would expect star formation to occur (e.g., cool-core clusters), the actual fraction of \textit{all} low redshift BCGs that possess star formation is unknown, as is the significance of the star formation. Is it great enough to add an appreciable amount to the overall mass of the galaxy?
Is Perseus A unique in its high star formation, or is it just a phase all BCGs pass through? To examine these questions, a complete, local sample of BCGs is required. 

%ORGANIZATION AND COSMOLOGY
This letter quantifies the fraction of star forming BCGs by creating an X-ray luminosity limited catalogue of local BCGs (Section~\ref{BCGSample}), 
measuring the IR photometry using the recently released AllWISE survey in Section~\ref{Photometry}, and using the 12 $\mu$m band to estimate the SFRs and specific star formation rates (sSFR) 
in Section~\ref{SFandBCGgrowth}. 
The cosmology used throughout this paper is $H_{0}=70 \textrm{km}~\textrm{s}^{-1}~\textrm{Mpc}^{-1}$, $h_{0} = \textrm{H}_{0}/100$, $\Omega_{M}=0.3$ and $\Omega_{\Lambda}=0.7$. 
All magnitudes are in the Vega system.

\section{BCG Sample}
\label{BCGSample}
A homogeneous BCG sample was created based on X-ray selection of host clusters with $L_{x}>1\times10^{44}$erg $\textrm{s}^{-1}$ in the \textsc{ROSAT} 0.1--2.4 keV band, corresponding to an approximate cluster mass of $M_{2500}\gtrsim 1 \times 10^{14}\textrm{M}_{\odot}$ \citep{Hoekstra11}. 
Selecting only X-ray luminous clusters ensures BCGs are located in comparably massive clusters. 
X-ray clusters were taken from the ROSAT Brightest Cluster Survey \citep[BCS;][]{Ebeling98}, extended BCS \citep{Ebeling00}, the X-ray Brightest Abell Clusters Survey \citep{Ebeling96}, 
the ROSAT North Ecliptic Pole Survey \citep[e.g.,][]{Gioia01}, the ROSAT-ESO flux limited X-ray galaxy cluster survey \citep[e.g.,][]{Bohringer01} and The Northern ROSAT All-Sky 
Galaxy Cluster Survey \citep{Bohringer00},
queried 
using the Base de Donn\'{e}es Amas de Galaxies \citep[BAX;][]{Sadat04}. 
The redshift of the sample was limited to $z<$ 0.1, creating a sample of 267 nearby, X-ray bright clusters where the completeness limit of each of the input surveys was at least 80\%. As we wished to identify BCGs regardless of their SFR, no colour selection criteria were applied.

We identified 144 BCGs by cross-matching the cluster sample with the BCG catalogues of \citet{Stott08}, \citet{Coziol09} and \citet{Wen12}.
 The remaining 123 unmatched clusters were inspected visually by AFM and KAP in both optical (Digitized Sky Survey) and infrared (Two Micron All Sky Survey, 2MASS) images, along with a 
NASA/IPAC Extragalactic Database object search to verify the redshifts of the candidate BCGs. The vast majority of identifications were unambiguous, and in all cases, the brightest galaxy in the 2MASS K-band at the cluster redshift was chosen as the BCG.
 
In some cases, clusters had to be eliminated from the sample if it was not clear which galaxy was the BCG. 
Reasons for this included mergers of galaxies of similar magnitude (e.g., Abell 3825), source confusion (e.g., Abell 523), or unavailable redshift information for any 
candidate BCG (e.g., Abell 72). 
Clusters were also eliminated from the sample if the WISE image showed source contamination from nearby saturated stars, or in six cases, the same cluster was entered twice 
into BAX under different names.

The final sample comprises 245 BCGs, covering both the northern and southern hemispheres, for which source photometry was extracted for from 
the AllWISE Data Release Catalog. 

% PHOTOMETRY
\section{Photometry}
\label{Photometry}
WISE \citep{Wright10} completed a whole-sky survey in four IR bands: 3.4, 4.6, 12 and 22 $\mu$m (denoted bands W1-W4). 
Bands W1 and W2 coincide with the Rayleigh-Jeans tail of the stellar emission of a galaxy and provide an accurate estimate of stellar mass. 
The W3 band is sensitive to warm dust and polycyclic aromatic hydrocarbon (PAH) emission associated with HII regions and molecular clouds.

Photometry for each BCG in the sample was extracted in all bands from the newly released AllWISE Source Catalog, combining improved multi epoch photometry from the WISE and NeoWISE surveys. 
We use the recommended approach for extended sources from the AllWISE explanatory statement\footnote{http://wise2.ipac.caltech.edu/docs/release/allwise/faq.html}, briefly described below.

Sources that are resolved and associated with a 2MASS Extended Source Catalog (XSC) object (232 galaxies in our sample) are measured with elliptical apertures made using apertures scaled from 2MASS XSC apertures to account for the WISE PSF. 
AllWISE sources that are associated with a 2MASS XSC object but are not resolved in WISE (7 galaxies) are modelled with profile fit photometry to avoid overestimating flux. 
For the four sources that are extended but not associated with a 2MASS XSC object, a fixed aperture size of $16.5^{\prime \prime}$ based on curve of growth was used with the prescribed aperture correction 
applied to each band. 
Where the signal to noise ratio 
is less than 2 in the W3 band, the magnitude of the 95\% confidence brightness upper limit is instead used for analysis. 

% SECTION 4 - SFRS AND BCG GROWTH
\section{Star Formation Rates and BCG Growth}
\label{SFandBCGgrowth}
\begin{figure}
 \centerline{ 
\psfig{file=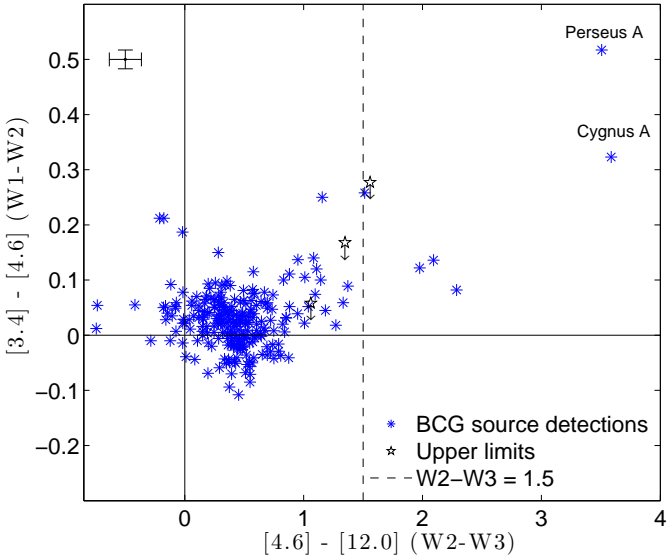,angle=0,width=3.5in}
}
\caption{WISE colour-colour diagram for BCGs in the sample, and representative error bars 
in the top left corner. Galaxies with [4.6]-[12.0]$>$1.5  
have an excess of IR emission in the W3 band, and are most likely to be dominated by star formation \citep{Cluver14}. 
The bulk of the sample lie offset from the zero point by $\sim0.4~\textrm{mag}$, indicating they are redder than the typical Rayleigh-Jeans region, in line with 
\citet{Jarrett11}. Two of the galaxies in the BCG sample are not detected in the W3 band, or did not have corresponding elliptical magnitudes in the case of the extended sources, so are not included here.}
\label{colorcolor}
\end{figure}
%DESCRIBE FIGURE 1
To illustrate the IR excess of BCGs, in Figure~\ref{colorcolor} we plot the WISE colours of BCGs with W3 band detections. 
Like local ellipticals \citep{Jarrett11}, the locus of the BCG sample is slightly offset from the origin at around W2-W3$\sim0.4$ (corresponding to an approximately Rayleigh-Jeans spectrum), with seven outlying BCGs residing in the star forming region of the plot. The 1$\sigma$ photometric scatter is $\sim$0.3 mag, and we expect 45 galaxies redder than this threshold, more than the amount of blue galaxies, showing this colour distribution is not the result of photometric scatter.
Just $3\pm1\%$ of the sample have an IR excess measured by a W2-W3 colour greater than 1.5, corresponding to systems likely dominated by star formation \citep{Cluver14}.

\begin{figure}
 \centerline{ 
\psfig{file=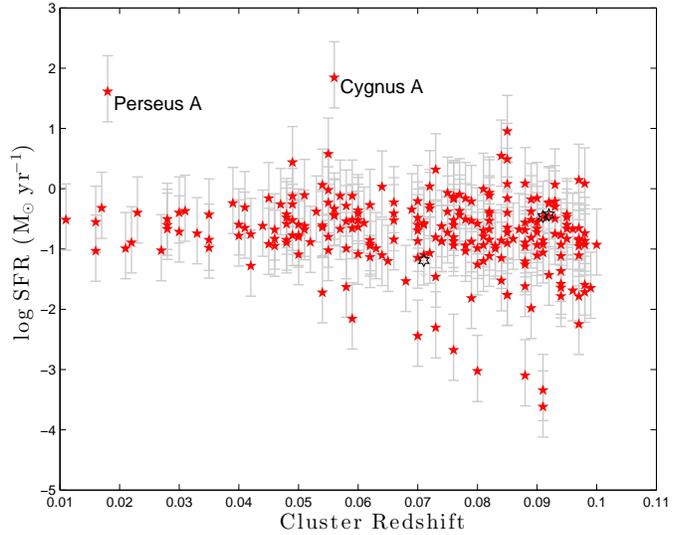,angle=0,width=3.5in}
}
\caption{BCG sample SFRs as calculated by the relation in \citet{Cluver14} as a function of cluster redshift with uncertainties derived from the scatter in Equation~\ref{Eq1} (Cluver 2014, private communication). Stellar emission has been subtracted from the W3 measurements to leave predominantly 
emission from hot dust.
Unfilled markers with arrows represent upper limits. Star formation rates are less than $10~\textrm{M}_{\odot}~\textrm{yr}^{-1}$ for $99\pm0.6\%$ of the sample, with the main exceptions of Perseus A 
and Cygnus A, exhibiting high star formation rates in comparison to the rest of the sample.}

\label{SFR}
\end{figure}
 %THE W3? SFR EQN
A H$\alpha$-derived SFR relation was determined for the WISE W3 band by \citet{Cluver14} by cross-matching WISE sources with those with available SFRs from GAMA I \citep{Gunawardhana13}: 

\begin{equation}
\log~\textrm{SFR}_{H\alpha}(\textrm{M}_{\odot}~\textrm{yr}^{-1}) = 1.13\log~\nu L_{W3}(L_{\odot})-10.24.
 \label{Eq1}
\end{equation}

We subtract stellar emission using the \citet{Cohen92} stellar calibration model fitted to the W2 band, which closely resembles the tail of the Rayleigh Jeans spectrum. Subtracting this primarily stellar emission from the W3 band leaves emission solely from warm dust.  Hence $L_{W3}$ is the stellar emission-subtracted IR luminosity in the WISE W3 band, and $\nu$ the effective frequency of the W3 passband. 

%K-CORRECTIONS
We modelled the WISE apparent and absolute magnitudes of galaxies as a function of both redshift and observed colour using the SED templates of \citet{Brown13}. 
The k-corrections as a function of colour at a given redshift were estimated and applied.
K-corrections are -0.3 magnitudes or less for the bulk of the galaxies in our sample, corresponding to a $\Delta \textrm{SFR}\sim0.04~\textrm{M}_{\odot}~\textrm{yr}^{-1} $.

%DESCRIBE FIGURE 2 Ð OUTLINE PER A AND CYG A OUTLIERS
The SFRs as calculated by Equation~\ref{Eq1} are plotted against cluster redshift in Figure~\ref{SFR}. Of the BCG sample with detections in the W3 band, $99 \pm 0.6\%$ have SFRs of $10~\textrm{M}_{\odot}~\textrm{yr}^{-1}$ or less. Two 
significant outliers, Perseus A and Cygnus A, have SFRs of 41 and 70 $\textrm{{M}}_{\odot}~\textrm{yr}^{-1}$ respectively. 
BCG photometry, along with stellar mass and SFR estimates are listed in Table~\ref{datatable}.
The dominant uncertainty is that from the scatter from the relation of Equation~\ref{Eq1} (Cluver 2014, private communication).

We matched our BCG sample to the total IR and UV-derived SFRs of \citet{Hoffer12}, and the optically derived SFRs of \citet{Crawford99}, with 18, 49 and 6 matches to our sample respectively.
For SFRs $>10~\textrm{M}_{\odot}~\textrm{yr}^{-1}$, our measured SFRs agree with \citet{Hoffer12} IR SFR within 26\%. IR SED fits are available for Perseus A and Cygnus A \citep[e.g.,][]{Mittal12,Privon12}, providing SFRs of $24\pm1$ and $\sim10 \textrm{M}_{\odot}~\textrm{yr}^{-1}$ respectively. These SFRs are lower than those derived for this study (41 and 70 $\textrm{{M}}_{\odot}~\textrm{yr}^{-1}$ for Perseus A and Cygnus A respectively), and for \citet{Hoffer12}'s total IR and UV derived SFRs.

% ATTEMPT TO DESCRIBE NEW FIGURE 3 - SSFR AS A FUNCTION OF L_X
  
There is considerable scatter between different SFR estimates for BCGs with SFRs $<10~\textrm{M}_{\odot}~\textrm{yr}^{-1}$, and the \citet{Cluver14} relation may underestimate SFRs for low IR luminosity galaxies. In the low redshift ($z<0.05$) regime, the \citet{Cluver14} relation works well for galaxies with SFRs $>3\textrm{M}_{\odot}~\textrm{yr}^{-1}$, but may underestimate SFRs lower than this. Despite this, for SFRs $<3\textrm{M}_{\odot}~\textrm{yr}^{-1}$, our SFRs agree with the multi-wavelength-derived SFRs of \citet{Hoffer12} and \citet{Crawford99} within 1 dex. Importantly, the scatter introduced in the low SFR sample either by the SFR relation, or possible excess emission by AGN contamination \citep[e.g.,][]{Donley08,Donoso12}  is not enough to push any low SFR galaxies into the highly star forming regime, preserving our primary conclusion that only a tiny fraction ($<1\pm0.6\%$) of local Universe BCGs are highly star forming.

%STELLAR MASS ESTIMATE
BCG stellar masses were determined using the relation of \citet{Wen13} from the WISE W1 band: 
\begin{multline}
\label{stellarmass}
 \log~\left( \frac{M_{*}}{M_{\odot}} \right) =  (-0.040\pm0.001) + \\ (1.120\pm0.001) \log~\nu L_{W1} (L_{\odot}),
\end{multline}

\begin{figure}
 \centerline{ 
\psfig{file=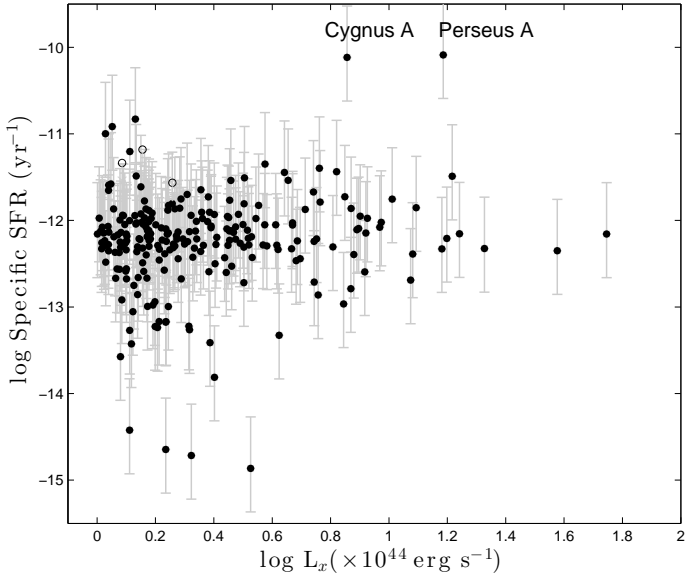,angle=0,width=3.6in}
}\caption{Specific star formation rates of BCGs in the sample as a function of cluster X-ray luminosity. Unfilled circles with arrows represent upper limits. There is little trend in sSFR with increasing $L_{x}$, and 
with the exception of Perseus A and Cygnus A, all BCGs at z$<$0.1 have sSFRs of less than 1\% per Gigayear.}
\label{sSFR}
\end{figure}

%DESCRIBE FIGURE 3 Ð MASS ADDED NOT SIGNIFICANT, COMPARISON OF RESULTS TO OTHER WORKS, AND FINAL STATEMENT Ð NOT SIG SF
\noindent{where $L_{W1}$ is the W1 band luminosity. Figure~\ref{sSFR} shows the sSFRs of the entire sample as a function of cluster X-ray luminosity. 
Most BCGs show low sSFR, of the order $10^{-12}~\textrm{yr}^{-1}$. 
Again, Perseus A and Cygnus A stand out with sSFR of almost $10^{-10}~\textrm{yr}^{-1}$, and while these SFRs will double the stellar mass of the galaxies over a Hubble time, they are unlikely to be sustained, as both systems are likely to be undergoing merger-induced star formation \citep[e.g.,][and references therein]{Holtzman92,Markevitch99,Conselice01}. }
Both Perseus A and Cygnus A are located in highly X-ray luminous clusters, suggesting there may be a slight preference for high $L_{X}$ clusters to host highly star forming BCGs \citep[e.g.,][]{ODea08}. Perseus A and Cygnus A are also both located in cool core clusters \citep[e.g.,][]{Hoffer12}.

Thus we conclude that while some star formation is occurring, this constitutes only modest BCG growth at current times.

\section{Summary}
We employed new mid-IR photometry from AllWISE to measure the SFRs of 245 BCGs in X-ray selected clusters at $z<0.1$. Our BCG catalogue was created by selecting X-ray bright clusters with $L_{x} > 1 \times 10^{44}~\textrm{erg}~\textrm{s}^{-1}$ in the \textsc{ROSAT} 0.1--2.4 keV band.
 
For the first time, 12 $\mu$m photometry (a SFR indicator) was measured for a large sample of local BCGs,  and we find that the majority have IR spectral energy distributions that differ from simple Rayleigh-Jean spectra. 
The bulk of BCGs at $z< 0.1$ possess little or no star formation at current times, with inferred SFRs of less than $10~\textrm{M}_{\odot}~\textrm{yr}^{-1}$ for  $99\pm0.6\%$  of local BCGs.
While the SFRs of BCGs with very low IR luminosities may have been underestimated, this does not impact our primary conclusion that only a tiny fraction of BCGs possess SFRs $>10~\textrm{M}_{\odot}~\textrm{yr}^{-1}$.
Hence we determine that only modest BCG growth is occurring as a result of star formation at the current epoch. 

Perseus A and Cygnus A are the BCGs with the highest SFRs in the $z<$ 0.1 Universe (41 and 70 $\textrm{M}_{\odot}~\textrm{yr}^{-1}$ respectively), calculated from hot dust emission. 
Whilst Perseus A is the archetypal example of AGN feedback, it is an exceptional BCG within the local Universe. 

\begin{table*}
\caption{WISE photometry and derived star formation rates of BCGs in the BAX X-ray selected cluster sample with $L_{x}>1\times 10^{44}~\textrm{erg}\textrm{s}^{-1}$. When there was no W3 detection for 
a BCG, the photometry and star formation rates are left blank. Similarly for when the BCG could not be unambiguously identified, 
we report the cluster coordinates and leave the stellar mass and star formation rates blank. Reasons for each BCG exclusion are listed in the footnotes. Table 1 is published in its entirety 
in the electronic edition.}
 \label{datatable}
 \begin{tabular}{L{2.5cm}C{1.5cm}C{1.5cm}C{0.75cm}C{0.95cm}C{1.7cm}C{1.7cm}C{1.7cm}C{1.5cm}C{1.5cm}}
  \hline

 Cluster       & BCG RA                   & BCG Dec                 & Cluster& Cluster& W1     & W2     & W3     & Stellar Mass                              & SFR                                                      \\
   Name       & (\textdegree, J2000) & (\textdegree, J2000) &   z       &        L$_{x}(10^{44}$             & (mag) & (mag) & (mag) & ($10^{10}\mathrm{M}_{\odot}$) & ($\mathrm{M}_{\odot}~\mathrm{/yr}$)  \\  
                   &                                   &                                  &             &$\textrm{erg/s}$)& & & &                                                             &                                                               \\
  \hline
  RXCJ0003.8+0203      &  0.9570     & 2.0665     & 0.092     & 1.50     & 11.51 $\pm$ 0.01     & 11.47 $\pm$ 0.01     & 11.07 $\pm$ 0.21     & 64.5 $ ^ { + 1.13 } _ { - 1.11 } $     & 0.35$^{+0.41}_{-0.09}$ \\[+.04in]
  RXCJ0011.3-2851      &  2.8403     & -28.8547     & 0.062     & 2.42     & 11.23 $\pm$ 0.01     & 11.21 $\pm$ 0.01     & 11.03 $\pm$ 0.16     & 34.1$^{+0.62}_{-0.61}$     & 0.07$^{+0.09}_{-0.02}$ \\ [+.04in]
  RXCJ0013.6-1930      &  3.3917     & -19.4838     & 0.094     & 2.26     & 12.19 $\pm$ 0.01     & 12.10 $\pm$ 0.01     & 11.84 $\pm$ 0.28     & 33.5$^{+0.58}_{-0.57}$      & 0.12$^{+0.14}_{-0.03}$ \\ [+.04in]
  RXCJ0017.5-3511      &  4.3958     & -35.1833     & 0.095     & 1.22     & 12.54 $\pm$ 0.01     & 12.49 $\pm$ 0.02     & 11.89 $\pm$ 0.20     & 24.0$^{+0.40}_{-0.39}$      & 0.22$^{+0.26}_{-0.06}$ \\ [+.04in]
  ABELL0021$^{a}$     &  5.1545     & 28.6590     & 0.095     & 2.64     & --     & --     & --     & --     & -- \\ [+0.04in]
  IVZW015      &  5.4033     & 28.0505     & 0.094     & 1.72     & 11.95 $\pm$ 0.01     & 11.89 $\pm$ 0.01     & 12.06 $\pm$ 0.36     & 43.3$^{+0.75}_{-0.73}$      & 0.02$^{+0.03}_{-0.01}$ \\ [+.04in]
  RXCJ0034.2-0204      &  8.5612     & -2.0846     & 0.082     & 2.40     & 11.37 $\pm$ 0.01     & 11.34 $\pm$ 0.01     & 10.49 $\pm$ 0.10     & 56.7$^{+1.01}_{-0.99}$      & 0.86$^{+1.02}_{-0.22}$ \\ [+0.04in]
  RXCJ0034.6-0208$^{b}$ & 8.6500  &  -2.1400  & -- & -- & -- & -- & --&--&--\\ 
  ABELL0072$^{b}$& 9.6366  &   45.7247  & -- & -- & -- & --& --&--&-- \\ [+.06in]
  ABELL0077      &  10.1180     & 29.5557     & 0.071     & 1.74     & 11.28 $\pm$ 0.01     & 11.26 $\pm$ 0.01     & 10.83 $\pm$ 0.13     & 44.5$^{+0.49}_{-0.48}$      & 0.25$^{+0.13}_{-0.03}$ \\ [+.04in]
%  ABELL0085      &  10.4602     & -9.3033     & 0.055     & 9.41     & 10.23 $\pm$ 0.01     & 10.23 $\pm$ 0.01     & 9.66 $\pm$ 0.10     & 72.2$^{+0.84}_{-0.83}$      & 0.59$^{+0.92}_{-0.20}$ \\ [+.04in]
%  ZWCL0040.8+2404      &  10.9674     & 24.4059     & 0.083     & 3.18     & 12.07 $\pm$ 0.01     & 12.01 $\pm$ 0.01     & 11.71 $\pm$ 0.19     & 28.4     & 0.11 \\ 
%  ABELL0104      &  12.4576     & 24.4505     & 0.082     & 2.03     & 11.54 $\pm$ 0.01     & 11.49 $\pm$ 0.01     & 10.61 $\pm$ 0.10     & 47.6     & 0.77 \\ 
%  RXCJ0051.3-2831      &  12.8519     & -28.4978     & 0.100     & 1.06     & 12.48 $\pm$ 0.01     & 12.34 $\pm$ 0.01     & 12.05 $\pm$ 0.29     & 28.8     & 0.12 \\ 
%  ABELL0119      &  14.0671     & -1.2555     & 0.044     & 3.30     & 10.20 $\pm$ 0.01     & 10.24 $\pm$ 0.01     & 9.80 $\pm$ 0.13     & 44.3     & 0.24 \\ 

  \hline
  \multicolumn{10}{l}{$^{a}$No extended source data in AllWISE catalogue} \\
  \multicolumn{10}{l}{$^{b}$Source Confusion} \\
 \end{tabular}

\end{table*}

\section{Acknowledgements}
The authors wish to thank the referee for thoughtful and insightful comments that have improved this manuscript, and Alain Blanchard, Michelle Cluver and Tom Jarrett for helpful conversations on the topic. 
AFM acknowledges support from an Australian Postgraduate Award (APA), and a J. L. William postgraduate award.
MJIB acknowledges financial support from the Australian Research Council 
(DP110102174, FT100100280) and the Monash Research Accelerator Program.\\
This research has made use of the X-Rays Clusters Database (BAX)
which is operated by the Laboratoire d'Astrophysique de Tarbes-Toulouse (LATT),
under contract with the Centre National d'Etudes Spatiales (CNES).\\
This publication makes use of data products from the Two Micron All Sky Survey, 
which is a joint project of the University of Massachusetts and the Infrared 
Processing and Analysis Center/California Institute of Technology, funded by 
the National Aeronautics and Space Administration and the National Science Foundation.\\
AllWISE makes use of data from WISE, which is a joint project of the University of California, 
Los Angeles, and the Jet Propulsion Laboratory/California Institute of Technology, and NEOWISE, 
which is a project of the Jet Propulsion Laboratory/California Institute of Technology. WISE and 
NEOWISE are funded by the National Aeronautics and Space Administration.\\
This research has made use of the NASA/IPAC Extragalactic Database (NED), which is operated by the Jet Propulsion Laboratory, California Institute of Technology, under contract with the National Aeronautics and Space Administration.

\bibliographystyle{mn2e}
\bibliography{BCGbib}

\end{document}